\documentclass[12pt,preprint]{aastex}

\newcommand{\be}{\begin{equation}}
\newcommand{\ee}{\end{equation}}
\newcommand{\beq}{\begin{eqnarray}}
\newcommand{\eeq}{\end{eqnarray}}

\shorttitle{The effect of the background on loop widths}
\shortauthors{L\'opez Fuentes, D\'emoulin \& Klimchuk}

\begin{document}

\title{Are constant loop widths an artifact of the background 
and the spatial resolution?}


\author{M. C. L\'opez Fuentes\altaffilmark{1,2,3}, 
        P. D\'emoulin\altaffilmark{4},        
        J. A. Klimchuk\altaffilmark{2}} 
\altaffiltext{1}{Instituto de Astronom\'{\i}a y F\'{\i}sica del Espacio,
CONICET-UBA, CC. 67, Suc. 28, 1428 Buenos Aires, Argentina} 
\altaffiltext{2}{Naval Research Laboratory, Code 7675, Washington, DC 20375}
\altaffiltext{3}{Member of the Carrera del Investigador Cient\'{\i}fico,
Consejo Nacional de Investigaciones Cient\'{\i}ficas y T\'ecnicas,
Argentina}
\altaffiltext{4}{Observatoire de Paris, LESIA, UMR 8109 (CNRS), F-92195, 
Meudon Principal Cedex, France}

\begin{abstract}

We study the effect of the coronal background in the
determination of the diameter of EUV loops, and we analyze the 
suitability of the procedure followed in a previous paper (L\'opez
Fuentes, Klimchuk \& D\'emoulin 2006) for characterizing their expansion 
properties. For the analysis we create different synthetic loops and 
we place them on real backgrounds from data obtained with the Transition 
Region and Coronal Explorer (\textit{TRACE}). We apply to these loops the
same procedure followed in our previous works, and we compare the results
with real loop observations. We demonstrate that the procedure allows us 
to distinguish constant width loops from loops that expand appreciably 
with height, as predicted by simple force-free field models. This holds 
even for loops near the resolution
limit. The procedure can easily determine when loops are below resolution 
limit and therefore not reliably measured. We find that small-scale 
variations in the measured loop width are likely due to imperfections 
in the background subtraction. The greatest errors occur in especially 
narrow loops and in places where the background is especially bright 
relative to the loop.  We stress, however, that these effects do not 
impact the ability to measure large-scale variations.  The result that 
observed loops do not expand systematically with height is robust. 

\end{abstract}



\section{Introduction}
\label{intro}


Since the solar corona is optically thin, studies based on 
coronal loop observations must include some form of subtraction 
of the background 
contribution (see e.g., Klimchuk 2000, Schmelz \& Martens 2006, 
L\'opez Fuentes, Klimchuk \& D\'emoulin 2006). In their recent 
statistical study based 
on observations from the Transition Region and Coronal Explorer 
(\textit{TRACE}, see Handy et al. 1999), Aschwanden, Nightingale \& 
Boerner (2007) showed that the 
background can be several times brighter than the loops themselves. It is 
likely that the background corona is formed by 
a number of loops that are too faint to produce a large enough 
contrast to make them detectable. However, these unobserved structures 
constitute a spatially fluctuating background for actual observed loops. 
Therefore, even for loops with constant intensity along their length, 
fluctuations due to the structuring of the background are expected.
The determination of morphological properties of a loop, 
such as its diameter, can be affected by the 
characteristics of the background, and therefore it is important 
that the background be taken into account during such analyses.


In a recent paper (L\'opez Fuentes, Klimchuk \& D\'emoulin 2006, 
henceforth LKD06) we explored the problem of the apparent constant
width of coronal loops. Since loops are the trace of magnetic flux 
tubes rooted in the photosphere, we might expect on the basis of 
simple force-free magnetic field models that most loops would expand 
with height. However,
observations show that this is not the case; both X-ray loops
(Klimchuk 2000) observed with the Soft X-ray Telescope (SXT, see 
Tsuneta et al. 1991) aboard Yohkoh, and EUV loops (Watko \& 
Klimchuk 2000, and LKD06) observed with \textit{TRACE}, seem to correspond 
to constant cross-sections.


In LKD06, we compared a number of observed \textit{TRACE} loops with 
corresponding model flux tubes obtained from force-free extrapolations of  
magnetogram data from the Michelson Doppler Imager (MDI, see Scherrer
et al. 1995) aboard the Solar and Heliospheric Observatory (\textit{SOHO}). To 
quantify the expansion of the loops and flux tubes, we defined the 
expansion factor $\Gamma$ as the ratio 
between the widths averaged over the middle and footpoint sections. 
We found that the mean expansion factor of the model  
flux tubes is about twice that of the corresponding observed loops. 
Another important result is that the cross section is much more asymmetric 
(from footpoint to footpoint) for the model flux tubes 
than for observed loops. We suggest that the origin of this asymmetry 
lies is the complexity 
of the magnetic connectivity of the solar atmosphere. In LKD06 we proposed 
a mechanism to explain the observed symmetry of real loops. 


Although the measured widths of observed loops have very little global 
variation, there are short distance fluctuations as 
large as 25\% of the average width. In LKD06 the loop background was 
subtracted by 
linearly interpolating between the intensities on either side of the loop. 
Since the background 
intensity can be as much as three to five times the intrinsic intensity 
of the loop, we might expect the width determination to be less reliable 
at positions where the ratio of loop to background intensities is smaller. 
If so, then measured width variations might be partly or largely an artifact 
of imperfect background subtraction. 


It is also possible that the width fluctuations are indicative of real 
structural properties of the loops. For 
instance, loops may be bundles of thinner unresolved 
magnetic strands that wrap around each other. If there are only a few 
such strands, then we might expect the width of the bundle to fluctuate on 
top of a global trend. Furthermore, 
the width should be anti-correlated with the intensity, since the bundle 
will be thinner and brighter in places where the strands 
are lined up along the line of sight, and it will be fatter and fainter 
in places where the stands are side-by-side across the plane of the sky. 


The filamentary nature of coronal loops, and the solar corona in general, 
has been progressively evident from the combination of models and 
observations (for a review, see e.g., Klimchuk 2006).
Our ability to discern the internal structure of loops is limited by
the instrument resolution. It can be seen from \textit{TRACE} images that
structures many times wider than the instrument Point Spread Function
(PSF) are clearly made of thinner strands. It is not surprising then that 
recognizable ``individual'' loops are no thicker than a few times the 
PSF width. 
Since identifiable individual loops are close to the resolution limit, 
it has been suggested that the apparent constant width may just be an 
artifact of the resolution (see the recent paper by DeForest 2007).  If loops 
are everywhere much smaller than the PSF, then they will appear to have 
a constant width equal to that of the PSF, even if the true width is 
varying greatly.  We have carefully accounted for the PSF in our earlier 
studies and concluded that this is not a viable explanation for the observed 
constant widths.  What we have not 
addressed in as much detail is the possible role of imperfect background 
subtraction.  This paper describes a study that addresses both the 
background subtraction and finite resolution and the extent to which 
they influence the measured widths of loops.


Our approach is to produce synthetic loops 
with constant and variable cross-sections, and place them on real 
\textit{TRACE} backgrounds to simulate loop observations. We then process the 
synthetic data following the same procedure used in LKD06, so we
can compare them with actual \textit{TRACE} loops. This allows us to 
determine whether the procedure followed in LKD06 is able to 
distinguish expanding loops from constant cross-section loops.   
We will answer the question of whether the lack of global expansion 
in observed loops is real or simply an observational artifact, as suggest 
by DeForest (2007). We will also investigate the reliability of the 
shorter length scale fluctuations that are often observed.


In Section~\ref{data} we describe the main properties of the
set of loops studied in LKD06. We explain the synthetic loop 
construction in Section~\ref{synthloops}. In Section~\ref{results} 
we compare synthetic and observed loops, and we discuss and conclude 
in Section~\ref{discussion}. 


\section{Observed \textit{TRACE} loops}
\label{data}


In LKD06 we studied a set of 20 loops from \textit{TRACE} images
in the 171~\AA~ passband. To determine the width of the loops we 
followed a procedure based on the measurement of the second moment
(the standard deviation) of the cross-axis intensity profile at each position 
along the loop. The measurement is done on a straightened version 
of the loop, as described in LKD06. Assuming circular cross-sections 
and uniform emissivity, the cross-section diameter (that we refer to as 
the width) will be 4 times the standard deviation of the profile.
The same procedure had been used in previous studies (see Klimchuk et 
al. 1992, Klimchuk 2000, Watko \& Klimchuk 2000). The actual background is 
estimated by linear interpolation of the background pixels at both 
sides of the loop. The obtained width is corrected for instrumental 
resolution (i.e. the combined PSF due to telescope smearing and detector 
pixelation). 

\begin{figure*}   
   \centering
  \hspace{0cm}
\includegraphics[bb= 105 250 465 610,width=12cm]{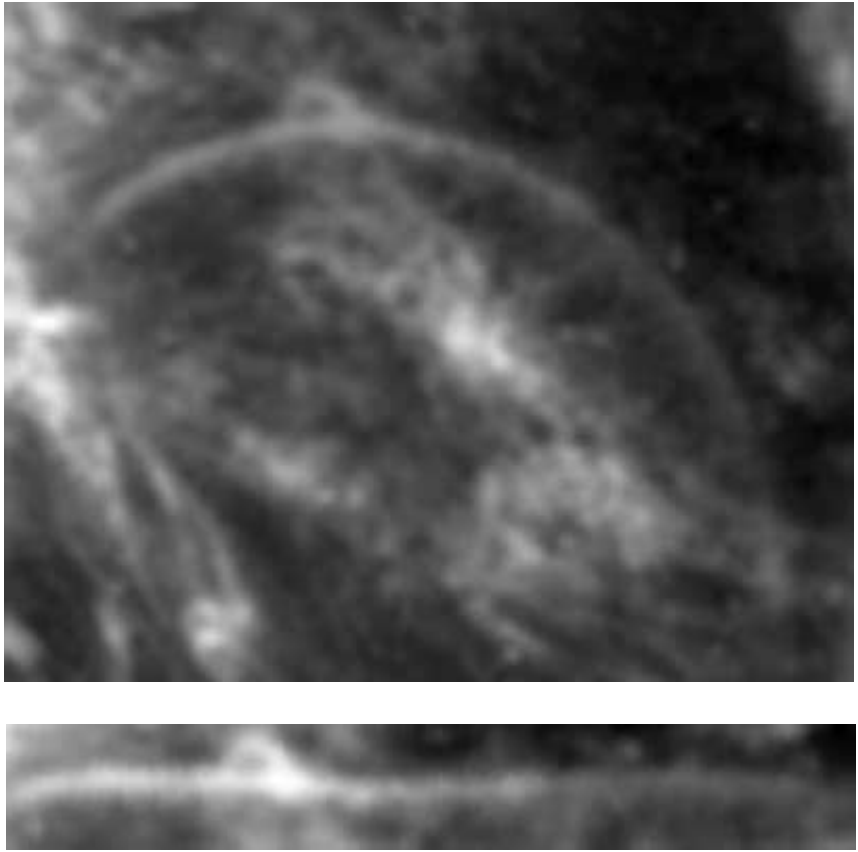}
      \caption{The loop shown is an example of the \textit{TRACE}
loops studied in LKD06. The lower panel shows the straightened 
version of the loop that is used for the width determination.}
         \label{f1}
\end{figure*}

The typical length of the studied loops is around 150 \textit{TRACE} 
pixels or 54 Mm, though loops as long as 300 pixels (108 Mm) are included in
the set. The average width for all loops in the set is 4.2 pixels or 1.5 Mm.
Figure~\ref{f1} shows a typical case having the average width and length. 
The upper panel shows the loop as observed in the \textit{TRACE} image, and 
the lower panel is the ``straightened'' version.


For the resolution correction we use a conversion curve (see Figure 4
in LKD06) to transform each measured standard deviation value to width. 
The curve has been obtained assuming a Gaussian PSF with a full width 
at half maximum of 2.25 pixels and loop cross-sections that are circular 
and uniformly filled. 
The chosen PSF width is an upper limit for values obtained in different 
studies (Golub et al. 1999, Gburek, Sylwester \& Martens 2006).  
The resolution correction curve plays a two-fold role. 
First, it allows us to obtain a more realistic width value from a measured
quantity like the standard deviation. Second, it is a filter
for measurements that are clearly unreliable.  
Standard deviation measurements smaller than a minimum value equal to the 
standard deviation of the PSF itself (where the conversion curve crosses the 
abscissa axis, see LKD06 Figure 4) are considered untrustworthy, and the
corresponding width is set to zero (i.e., rejected). 
Problems resulting from significant errors in the background 
subtraction can also be identified in this way. It is worth remarking that 
our approach is quite cautious in that the PSF we have assumed is wider than 
the most recent estimates (see Gburek, Sylwester \& Martens 2006). Some 
of the measurements we reject as being unresolved may in fact be valid.


Figure~\ref{f2} is a plot of width (asterisks) versus position along the 
loop shown in Figure~\ref{f1}. The horizontal line corresponds
to the average width. It is nearly identical to the mean width
of all the loops in the set. The three ``zero width'' values that 
lie on the abscissa axis correspond to standard deviation measurements that
were below the resolution limit as explained above.


To quantify the expansion of loops from footpoint to top 
we defined expansion factors as follows:

\be
\label{gamma}
\Gamma_{m/se} = \frac{2~W_{m}}{W_{s}+W_{e}},~~
\Gamma_{m/s} = \frac{W_{m}}{W_{s}}~~\textrm{and}~~
\Gamma_{m/e} = \frac{W_{m}}{W_{e}},
\ee

\noindent where $W_{m}$, $W_{s}$ and $W_{e}$ are the average width of 
portions that cover 15\% of the loop length at the middle, 
start footpoint, and end footpoint, respectively. Start and end refer to the 
magnetic field line traces used to define the magnetic flux tubes in  
the extrapolation models. The model flux tubes expand much more than the 
corresponding observed loops (LKD06). Their expansion 
factors are 1.5 to 2 times larger.


As explained in Section~\ref{intro}, the loop width fluctuates 
as much as 25\% over short distances (see e.g., Figure~\ref{f2}). 
We tried alternate measures of the loop width (full width at half 
maximum and equivalent width of the intensity profile), and the 
same fluctuations are present.
Our conclusion is that the fluctuations 
are most likely 
due to the influence of the background (see below). Since these fluctuations 
have a short length scale and vary quasi randomly around a global trend, 
they do not significantly affect 
the measured expansion factors. 


\section{Synthetic loops}
\label{synthloops}


In this study we create a set of synthetic loops with similar
characteristics to the \textit{TRACE} loops studied in LKD06, and we 
overlay them on real \textit{TRACE} backgrounds. The axis of the loop
is linear and its cross-section is circular. To analyze the possibility
that the apparent constant width is due to a resolution effect
we create loops of two kinds: loops with constant diameter along 
their length, and loops that are wider in the middle than at the 
footpoints. For the second class of synthetic loops we use an 
expansion factor that is typical of the model  
flux tubes obtained in LKD06. We set the diameter of the loop 
at the mid point to be twice the diameter at the ends, and  
we assume that the diameter varies quadratically with 
position.  Since the expansion factor 
defined in Equation~(\ref{gamma}) involves averages along 15\% 
sections of the loop, $\Gamma_{m/se} = 1.57$ rather than 2.0. 


We have chosen two kinds of background for the synthetic loops.
Background I, shown in the top-left panel  
in Figure~\ref{f3}, corresponds to a typical \textit{TRACE} loop background: 
it has similar intensity magnitude and fluctuations, and it contains moss (see 
Berger et al. 1999, Martens et al. 2000) and other intense features. 
Background II, shown in the top-right panel, is fainter and fluctuates less 
than Background I.
Although it does not correspond very well to real loop backgrounds,
we consider it interesting to study how this kind of background affects the
width determination. The average intensities
of backgrounds I and II are approximately 70 and 30~DN (Data Numbers), 
respectively. For comparison, the typical intrinsic (background subtracted) 
intensity of observed loops is between 20 and 40~DN.
Both background areas have been extracted from a \textit{TRACE} image
in the 171~\AA~ band obtained at 01:45 UT on July 30, 2002.


To create a simulated \textit{TRACE} image containing the synthetic loop 
we proceed as follows. We create an image of the loop without background. 
The maximum intensity of the loop (at the axis) is set proportional to 
the average intensity of the background image on which it will be later 
superposed. The constant of proportionality is referred to as the 
intensity factor $\Phi$. Since the background intensity tends to be 
higher than the intrinsic loop intensity, $\Phi$ is generally smaller 
than 1. To simulate the finite resolution, we smooth the image of the loop 
using a gaussian profile with a full width at half maximum of 2.25 
pixels corresponding to the instrument PSF. The resulting 
loop is then placed on the previously selected background (I or II) 
from the \textit{TRACE} image.


The images in Figure~\ref{f3} have been created as described above. 
Both panels in each row use the same synthetic
loop placed in one case on background I (left) and the other case on 
Background II (right).  The four loops differ in the following ways. 
The loop in the second row (panels a and b) has a constant 
diameter of 4 pixels, corresponding to $\Gamma_{m/se} = 1$. The loop in 
the third row (panels c and d) has a  
diameter that expands from 2.5 pixels at the ends to 5 pixels at the 
center, corresponding to $\Gamma_{m/se} = 1.57$. The loop in fourth row 
(panels e and f) has a constant diameter of
3 pixels. Finally, the loop in the bottom row (panels g and h) expands 
from 2 pixels at the ends to 4 pixels in the middle. 
Notice that the ends of this last loop are narrower than the PSF. 
The intensity factor $\Phi$ has been adjusted so that the 
resulting loops look similar, by eye, to typical \textit{TRACE} 
loops. We used $\Phi = 0.5$ for Background I and $\Phi = 0.7$ for 
Background II.   
Considering the average intensities of Backgrounds I and II, this
gives intrinsic loop intensities of around 35~DN and
25~DN, respectively.  These values are consistent with the intrinsic
intensities of observed loops. 


The photon statistical noise associated with \textit{TRACE} 
data is given by $\sqrt{N}$, where $N$ is the 
number of photon counts per pixel (Handy et al. 1999). Since 1~DN 
corresponds to 12 photon counts, the photon noise as a percentage 
of the signal is:

\be
\label{noise}
PN_{\%} = 100 \frac{\sqrt{12 I}}{12 I} \approx \frac{30}{\sqrt{I}},    
\ee 

\noindent
where $I$ is the intensity of the signal in DN/pix. 
The synthetic loop data constructed here includes the photon noise present 
in the \textit{TRACE} image used for the background.  As we now 
demonstrate, this contribution dominates the noise from the loop itself, so 
we can safely ignore the loop contribution.
Let us first consider the extreme case of low background and loop intensities,
namely:  $I_{b}=30$~DN and $I_{l}=10$~DN. According to 
Equation~(\ref{noise})  
the photon noise of the total signal is $PN_{\%} = 30/\sqrt{40}$ 
or 4.7\%. On the other hand, for our synthetic images 
(photon noise from the background only) it is 
$PN_{\%} = 30/\sqrt{30}$ or 4.1\%, meaning a difference of 0.6\%.
For a more typical case of $I_{b}=70$~DN and $I_{l}=25$~DN, the
same percentages are 3.1\% and 2.6\% respectively, implying a difference of
0.5\% of the total signal. These differences are minor 
and will have a negligible effect on the results of the following 
sections. 


\section{Results}
\label{results}

\subsection{Can we detect expanding loops?}
\label{results-1}


From the set of loop images, we measured the 
width following the same procedure used in LKD06 for real loops and 
described in Section~\ref{data}. We used the conversion curve (Figure 
4 in LKD06) to correct for the instrument PSF.
The non-linearity of the curve increases the dispersion of the resulting   
widths at smaller values approaching the width of the PSF. 
In Figure~\ref{f4} we plot the ``measured'' width (asterisks) versus
position along the loop for the eight cases in 
Figure~\ref{f3}. The format of the figures is the same. For comparison, 
we also plot as continuous lines the actual diameters used to construct 
the images. It can be seen 
that, despite the fluctuations, expanding and constant width loops
are clearly distinguishable. This is true for loops that are 
relatively wide (top two rows) and loops that are relatively narrow 
(bottom two rows). This demonstrates convincingly that, if loops 
expanded as expected from standard force-free extrapolation
models, then it would be noticeable from observations even when 
they are very close to the resolution limit (last row).  
Since that it not the case, this may imply that actual magnetic fields 
have more complexity than is present in the standard models. 
We know, for example, that the field 
is comprised of many thin flux strands (elemental kilogauss tubes) that 
are tangled by photospheric convection. We believe this can explain the 
symmetry of observed loops with respect to their summit 
(see discussion in LKD06 and Klimchuk 2006), but whether it can 
also explain the lack of a general expansion with height is unclear.


It is interesting to note from the plots in Figure~\ref{f4} that
the measured width is systematically smaller than the width set
in the construction of the loops.  This tendency appears in 
all loops and is very likely due to an underestimation of 
the real width in the measurement procedure. 
The procedure requires a subjective selection of the loop 
edges for the purpose of defining the background and computing the 
standard deviation.  
During this step one can miss the faint tail of the cross-axis 
intensity profile that blends in with the background.  
We have verified that there is a tendency to define the loop edges 
to be slightly inside the actual edges.  This causes 
the measured width to be artificially small, both because the 
tail of the profile is missing from the standard deviation computation and 
because too strong a background is subtracted from the loop.  We expect the 
effect to be greatest for loops that are especially faint or 
especially narrow, as discussed below.
If this explanation is correct, we can conclude  
that the \textit{TRACE} loops studied in LKD06 are actually slightly 
wider than our measurements seem to indicate. 

The fact that the measured width is a lower bound for the real width  
gives further support to our assertion that the analyzed loops are 
instrumentally resolved.  In LDK06, we estimated the width 
uncertainties associated with background subtraction by repeating 
each measurement using different choices for the loop edges.  We concluded that rule-of-thumb error bars range from 
10\% below to 20\% above the measured best value. It now appears that 
the actual error bars may be somewhat larger. However, we stress that 
this does not impact our ability to distinguish expanding loops from 
non-expanding loops, as is readily apparent from Figure~\ref{f4}.  


To quantify this claim, we computed the expansion factors ($\Gamma_{m/se}$ 
in Equation~\ref{gamma}) of all the synthetic loops shown in 
Figure~\ref{f4}.  These are listed in Table~\ref{table1}.
The upper and lower limits that define the error bars are the expansion 
factors $\Gamma_{m/s}$ and $\Gamma_{m/e}$. For comparison, in the case 
of the observed loop of Figures~\ref{f1} and~\ref{f2}, the
expansion factor computed in the same way is $1.03 \pm 0.04$.  
The values given in Table~\ref{table1} clearly confirm our conclusion that  
loops with constant and expanding cross section can be easily distinguished.
It is interesting to note that the expansion factors for the same
loops placed in different backgrounds can be notably different. The same 
is true for loops of the same kind (expanding or not) but with different 
characteristic size (wide or narrow). Compare
row 1 with row 3, and row 2 with row 4 in Figure~\ref{f4} and the table.
The error bars are also different
in all cases. Part of these differences may be due to the subjective
part of the analysis procedure (the selection of the loop edges). 
However, repeating the width measurements we obtain approximately the 
same expansion factors. Therefore, the distribution of the background 
emission and the characteristic size of the loop both play a role in 
determining the precise value of the expansion factors and the error 
bars. In particular, Backgrounds I and~II tend to give an under and over 
estimation of the expansion factors, respectively (compared to the 
values set during the loop construction). Nevertheless, we want to stress 
that the measured expansion factors of the expanding and non-expanding 
synthetic loops are clearly clustered around the actual values, implying that 
loops with constant and expanding cross section are readily distinguishable. 


Next, we study how the observed loop expansion is affected by
the relative intensity of the loop compared to the background. To test 
this, we created synthetic data in the way described in 
Section~\ref{synthloops}, for different values of the loop-to-background   
intensity ratio $\Phi$. 
In Figure~\ref{f5} we plot the expansion factor $\Gamma_{m/se}$
(Equation~\ref{gamma}) versus $\Phi$ for 4 narrow synthetic
loops with similar characteristics to those shown in panels e) to h) 
of Figures~\ref{f3} and~\ref{f4}. The difference is that the loops 
of Figure~\ref{f5} are 300 pixels long, instead of 200 pixels. 
The definition of $\Gamma$ is not affected by 
the change of length. On the other hand, longer loops provide 
more measurements and better statistics for studying how loop 
expansion depends on the loop-to-background intensity ratio. 
We chose thin loops for Figure~\ref{f5} because their expansion is 
more challenging to measure and they are more affected by the 
background.

If the intensity ratio $\Phi$ is too small, it is difficult to detect a loop 
above the background, much less measure its width. Our previous studies 
of observed loops have therefore avoided such cases. We subjectively 
define a lower limit for loop visibility of around $\Phi=0.3$. Below 
that, the width determination is unreliable. The upper value $\Phi=1.5$ 
is extreme for most \textit{TRACE} loops, but it is interesting for analysis and 
may be appropriate to other datasets. The intermediate $\Phi$ values 
are 0.5, 0.7 and 1.0. Figure~\ref{f5} provides strong additional support 
for our claim that the expansion factors of expanding and non-expanding 
loops can be clearly distinguished, even for the most critical cases of 
very low intensities and narrow widths. In no case do the error bars of 
expanding and non-expanding loops overlap.


It is interesting to note that the synthetic loops used for 
Figure~\ref{f5} overlap with more of the background image than do the 
shorter synthetic loops used for Figures~\ref{f3} 
and~\ref{f4}.  The footpoint and middle sections therefore combine
with different portions of the background.  Since the expansion factors 
are qualitatively similar in the corresponding cases, we can be confident 
that our results are not an artifact of the particular loop-background combinations.


So far we have not considered loops that are completely 
below the resolution limit.  In Figure~\ref{f6} we show two cases
of unresolved synthetic loops.  Both have a constant diameter of 0.5 pix, and 
both use Background I (Figure~\ref{f3}).  The loops differ only in the 
intensity ratio $\Phi$, which is set to 1 for case (a) and 3 for
case (b).  Note, however, that because the loops occupy only a fraction of a 
pixel, the ``observed'' intensity ratios are much smaller: around 0.25 for the
$\Phi$=1 loop and 0.5 for the $\Phi$=3 loop.  
 
Figure~\ref{f7} shows the widths of the loops as measured in 
the usual way, including correction for instrument resolution.  A majority of 
the measurements are equal to zero, meaning that the computed 
standard deviation is below that of the PSF.  This is especially true for the 
fainter loop of case (a).  We can understand this behavior as follows.  Due to 
the influence 
of the variable background, we expect some measurements to be too large and 
others to be too small.  However, because of the systematic effects associated 
with loop edge selection, discussed above, we expect more of the 
measurements to be too small.  

The conversion from standard deviation to width is very sensitive at 
small values, where the conversion curve is nonlinear, and it only 
takes small errors in the standard deviation to produce a zero width value. 
The solid line in Figure~\ref{f7} indicates the actual loop width of 0.5 pix, 
while the dashed line indicates the full width at half maximum of the PSF.  
The most important conclusion to draw from the figure is that our 
measurement technique can easily detect when 
loops are unresolved, i.e., when they are thinner than the PSF.  
As we stated before, the loops analyzed in LKD06 and previous works are 
all wider than the PSF (see also Section~\ref{discussion} below).


Finally, in Figure~\ref{f8} we plot width versus position along a 
synthetic loop with a footpoint width of 2.5 pixels and a model 
expansion factor of $\Gamma_{m/se}=2.2$. Our measurement procedure 
tracks the loop expansion very well. The expansion factor computed 
from the observed width as in Table~\ref{table1} gives 
$\Gamma_{m/se}=2.1 \pm 0.2$. Therefore, the loop can be readily 
distinguished from the $\Gamma_{m/se}=1.57$ loop having the same
footpoint size in Figure~\ref{f4}, panel (d). 

\subsection{Short length scale width fluctuations}
\label{results-2}

\begin{figure*}   
   \centering
  \hspace{0cm}
\includegraphics[bb= 95 370 550 720,width=15cm]{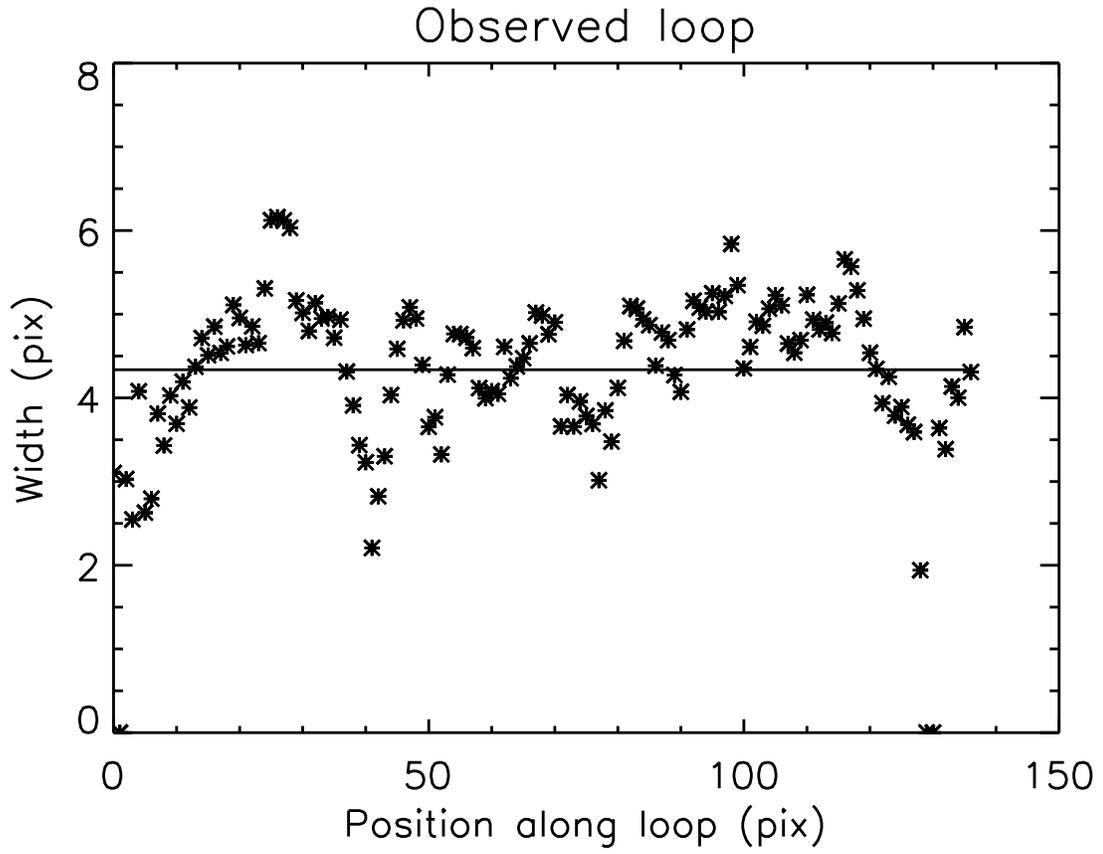}
      \caption{Measured width versus position along the
loop of Figure 1.  Background subtraction and PSF correction have been 
applied, as described in Section~\ref{data}. The horizontal line shows 
the mean value.}
         \label{f2}

\end{figure*}

As discussed in Section~\ref{intro} the measured widths of observed 
loops fluctuate as much as 25\% over short length scales. 
It is important to know whether these variations are real or an 
artifact of the background.  Comparison of panels a) and b) in 
Figure~\ref{f4} with Figure~\ref{f2} shows  
that synthetic and observed loops with similar characteristic width 
exhibit similar width fluctuations. For the observed loop of
Figure~\ref{f2}, the amplitude of the fluctuations computed as the 
ratio of the standard deviation of the measured width to its average
is 18\%. The corresponding ratios for the synthetic loops of panels 
(a) and (b) in Figure~\ref{f4}, are 17\% and 25\%, respectively. 
This suggests that the fluctuations are not real and argues against 
loops being comprised of a small number of braided strands (the possibility 
that they are bundles of {\it many} tangled strands is not affected).  This is 
not a firm conclusion, however, since the fluctuations are somewhat more 
coherent for the observed loop than for the synthetic loop.  We return to 
this issue below.   
Narrower loops (e.g., panels (e) and (f) in Figure~\ref{f4})
show larger amplitude fluctuations (21\% and 38\%, respectively) mostly 
because of the non-linearity of the resolution correction curve (LKD06 
Figure 4), which exaggerates differences at smaller widths.


To study how the background fluctuations affect the width determination, 
we analyze the relationships between the width and the loop and background 
intensities. In Figure~\ref{f9} we plot as a function of position along the 
loop: the intensity of the background pixels at either side of the loop 
(from which the loop background is linearly interpolated; continuous lines), 
the loop width (dotted), the loop intensity (maximum intensity of the 
background-subtracted profile; dashed), and the absolute value of the 
difference between the two background pixel intensities (dot-dashed). 
The loop width is given in pixels and multiplied by 10 for easier 
comparison with the intensities. The upper panel corresponds
to the observed loop example of Figures~\ref{f1} and~\ref{f2}, and the 
lower panel correspond to the synthetic loop of Figures~\ref{f3} 
and~\ref{f4}, panels (a). 


Figure~\ref{f9} shows that our synthetic loop data share the main
qualitative characteristics of real loops. The fluctuations of the
background intensity and its difference at the sides of the loop, 
and the loop intrinsic intensity and its fluctuations, are similar 
in both cases.
There are obvious differences due to the spatial structure unique to each 
background that can easily be identified in the images. For example,
the bumps between 30 and 60 for the observed loop, and between positions 
0 and 40 and between 90 and 130 for the synthetic loop, can 
be traced to patches of enhanced emission in Figures~\ref{f1} and~\ref{f3}.
Another difference is the global variation in the intensity of the 
observed and synthetic loops.  The measured intensity of the synthetic
loop is nearly constant because the loop was constructed 
that way (small fluctuations come entirely from imperfect background 
subtraction).  
The measured intensity of the observed loop, on the other hand, tends 
to diminish systematically toward the right 
end. This is likely to be real and not an artifact of the background 
subtraction. Despite of these expected differences, the
comparison shows that the synthetic loops reproduce the main properties
of the observed cases.


We have suggested that small-scale fluctuations of the measured 
intensities and widths of loops are due to imperfect background 
subtraction. To further assess this, we look for statistical 
correlations between these quantities. In the upper panels of
Figure~\ref{f10} we plot width versus intensity 
for all positions along the observed and synthetic loops, respectively. 
We find that there is a small 
direct correlation between the width and intensity in both cases: 
wider sections of the loops tend to be brighter. 
The lines in the scatter plots are least-squares fits, which 
have the indicated slopes and intercepts.  
The correlation between width and intensity can be explained by the 
tendency, during the interactive analysis procedure, to 
miss the wings of the intensity profile and define the loop edges to be inside 
the true edges.  As described earlier, this causes an over estimation of the 
background intensity and produces artificially narrow loop widths and 
artificially faint loop intensities.  We expect the magnitude of this effect 
to vary depending on the brightness of the background relative to the loop.  
It will be stronger (i.e., the underestimates of width and intensity will 
be greater) when the background is relatively bright. This is confirmed 
in the second row of Figure 10. It shows an inverse correlation between 
the measured width and the background-to-loop intensity ratio for both 
the observed and synthetic loops. The background intensity used here is 
the average of the sloping background subtracted during the analysis 
(i.e., the average of the values on either side of the loop). Notice 
also that for the synthetic loop, the measured width tends to be smaller 
than the model width (4 pixels) when the relative intensity of the 
background is larger.

This effect is almost certainly responsible for the width-intensity 
correlation of the synthetic loop and seems a likely explanation for the 
observed loop, as well.  Whether it is strong enough to allow the 
possibility that loops are bundles of a few (3-5) intertwined strands 
is unclear. Recall that such loops would exhibit an inverse correlation 
between width and intensity if the measurements were perfect. Are 
measurement errors large enough to negate this inverse correlation 
and produce a small direct correlation, as observed?  Only more 
involved modeling can answer this question.


It seems plausible that cross-loop gradients in the background could also have 
an effect on the measured width.  Certainly small scale inhomogeneities are 
more difficult to subtract than a flat background.  
In the upper panels of Figure~\ref{f11},
we plot width versus the absolute value of the background intensity 
difference on the two sides of the loop.
No correlation is apparent for either the observed or synthetic 
loops. We confirmed a lack of correlation using a
non-parametric statistical analysis. We also find no correlation 
between the intrinsic loop intensity and the background intensity 
difference. 


The right bottom panel of Figure~\ref{f11} indicates how the known 
error in width measurement for the synthetic loop correlates with the 
background intensity gradient. The ordinate is the absolute 
value of the difference between the measured width and the width 
used during the loop construction. The abscissa is the absolute 
value of the background intensity difference on the two sides, 
normalized by the loop intensity.  The normalization is meant to 
compensate for the fact that background gradients should 
have a lesser impact on bright loops.  The left bottom panel of 
Figure~\ref{f11} is a corresponding scatter plot for the observed 
loop. Since the actual width is not known, the ordinate is replaced  
by the absolute value of the deviation of the measured width from 
its mean. In neither case is there a correlation, as confirmed by 
statistical analysis. We conclude that the magnitude of the 
background has a bigger effect on the width measurements than 
does the difference in the background on the two sides of the loop 
(the cross loop gradient).  


\section{Discussion and conclusion}
\label{discussion}


In this paper we study the effect of the background and the 
instrument PSF in the determination of the apparent width of
EUV coronal loops observed by \textit{TRACE}. Our main motivation
is to extend the results obtained in our previous work: L\'opez Fuentes,
Klimchuk \& D\'emoulin (2006; LKD06). There, we compared a set
of observed \textit{TRACE} loops with corresponding force-free model
flux-tubes, and we found that observed loops do not expand with height
as expected from the extrapolation model. Here, we construct artificial 
loops with expansion factors similar to those of the studied loops and 
the model flux-tubes, and we overlay them on real \textit{TRACE} 
backgrounds.
We repeat on these synthetic loops the same procedure followed in LKD06, 
and compare the results back with real loops. 
We find that even for loops close to the resolution limit the procedure
followed in LKD06 discerns expanding and non-expanding cross-sections. 
The method includes a resolution correction that 
identifies measurements that are below the resolution limit and therefore 
unreliable (see explanation
in Section~\ref{data}). We used a gaussian Point Spread 
Function (PSF) for the instrument with a FWHM of 2.25 pixels, 
which is an upper bound for values found by different authors 
(Golub et al. 1999, Gburek et al. 2006).


In a recent paper, DeForest (2007) has proposed the interesting idea
that most thin individual loops observed by \textit{TRACE} are actually 
extremely bright structures well under the resolution limit. In this
scenario, the loop apparent width would be given by the instrument 
PSF. In this way, loops may actually expand, but their size both at the 
top and the footpoints would be unresolved and would appear the same.
The motivations for this conjecture are the apparent
constant width of loops, and the observation that \textit{TRACE} loops 
have an intensity scale height that is considerably larger than expected 
for static equilibrium (Winebarger et al. 2003, Aschwanden et al. 2001) 
or steady flow (Patsourakos \& Klimchuk 2004). 
More precisely, for expanding loops that are everywhere unresolved, 
the density gradient present in the corona is larger than inferred from 
the observations under the assumption of constant cross section.
   
According to the above explanation, we should expect all individual
\textit{TRACE} loops to have 
a true width less than that of the PSF and an apparent width roughly equal 
to that of the PSF.   
However, observations do not support this.  The mean width 
of the loops studied in LKD06 is 4.2~pix after  
correction for 
the instrument resolution (see also Watko \& Klimchuk 2000).  
As shown in Figure ~\ref{f7} and discussed in Section~\ref{results-1}, 
our method can easily identify loops that are intrinsically more narrow 
than the PSF.  The loops selected for our studies are clearly not of this 
type.

As we discussed in Section~\ref{intro}, coronal structures that are many
times wider than the \textit{TRACE} PSF are observed to be formed by 
thinner individual loops. Therefore, there is an intermediate range of
widths -- let us say between one and three PSF widths -- for which 
the profiles produced by unresolved threads could overlap to form 
apparently wider loops. This, together with the effect of a fluctuating and
intense background, are the arguments provided by DeForest (2007) to
explain loops wider than the PSF.  A key point in this discussion 
is that unresolved neighbor threads might be expected to separate from 
each other with height for the same reasons that individual 
strands might be expected to expand with height (e.g., if the field 
behaves like {\it simple} force-free extrapolation models predict).  
In this respect, a structure
formed by diverging threads does not differ from the expanding loops
studied in Section~\ref{synthloops}. As we discussed there, the plots
in Figures~\ref{f4} and~\ref{f5} and the $\Gamma$ factors given in
Table~\ref{table1} show clearly that our procedure for the width 
determination would be able to detect the expansion if it 
existed, even for loops near the resolution limit. 

\begin{figure*}   
   \centering
  \hspace{0cm}
\includegraphics[bb= 50 142 535 669,width=15cm]{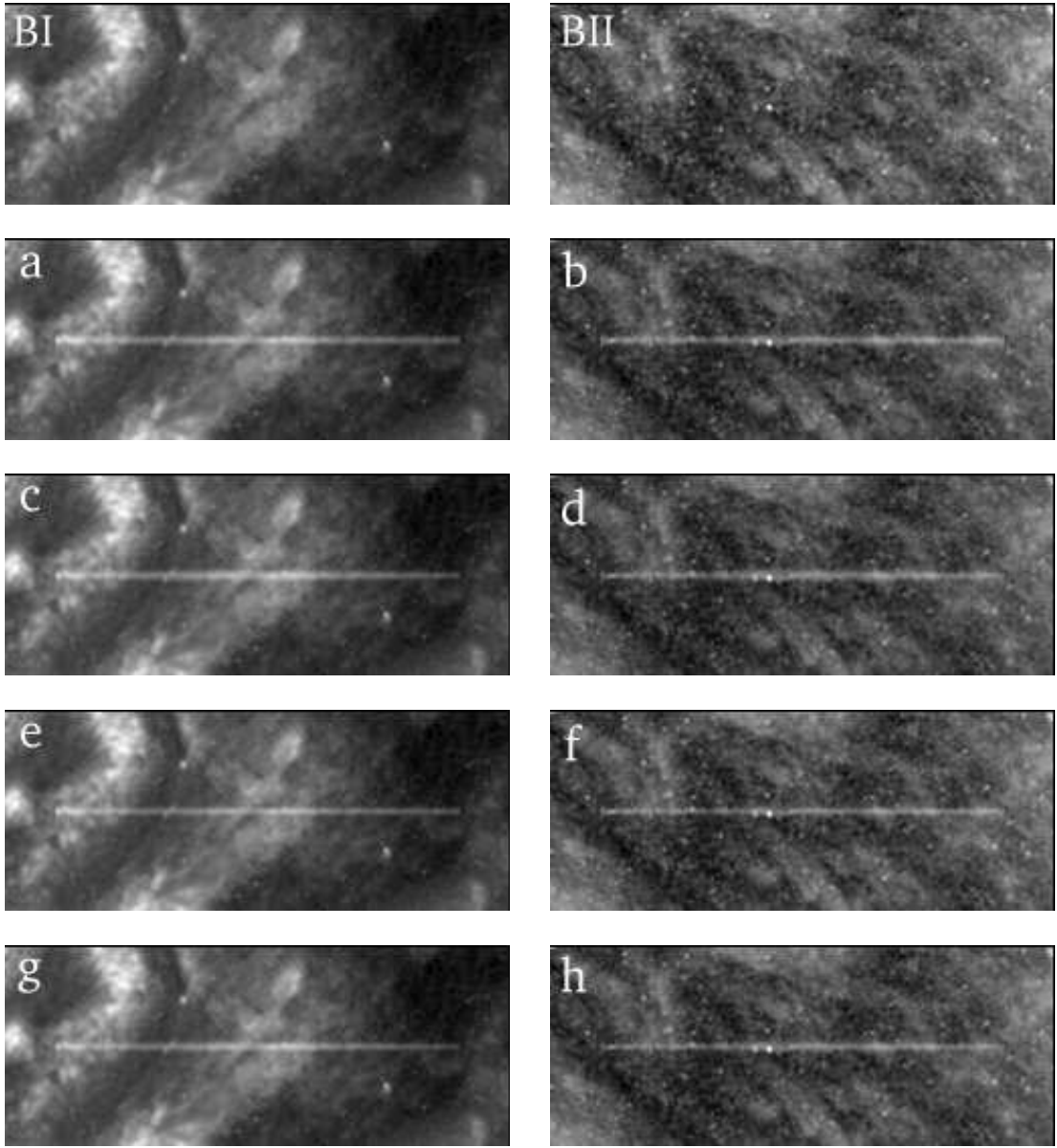}
      \caption{Synthetic data created by superposing loops with different
specified properties on real \textit{TRACE} backgrounds. 
The top panels show the background used in each column.
The ends (footpoints) and middle of the loops have an imposed
diameter (in pixels) of: (a,b) 4-4, (c,d) 2.5-5, (e,f) 3-3, (g,h) 2-4. 
This provides both constant and expanding (by a factor 2) synthetic 
loops close to the spatial resolution. For a detailed description
of the panels see Section~\ref{synthloops}.}
         \label{f3}
\end{figure*}

\begin{figure*}   
   \centering
  \hspace{0cm}
\includegraphics[bb= 60 70 535 790,width=15cm]{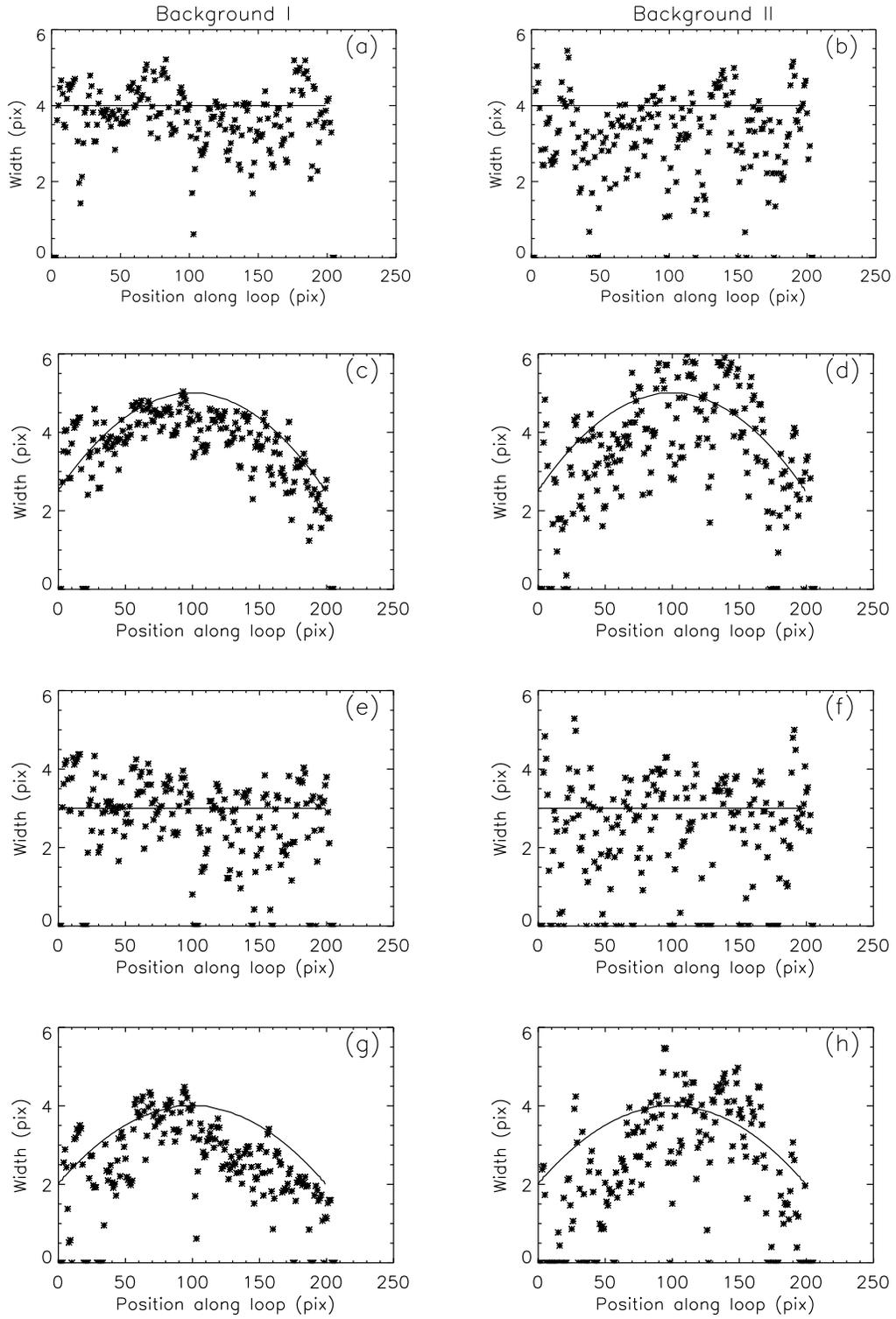}
      \caption{Measured width corrected for instrument resolution (asterisks) 
versus position along the synthetic loops shown in panels a) to 
h) of Figure~\ref{f3}. Continuous lines indicate the model width.}
         \label{f4}
\end{figure*}

\begin{figure*}   
   \centering
  \hspace{0cm}
\includegraphics[bb= 60 250 560 630,width=16cm]{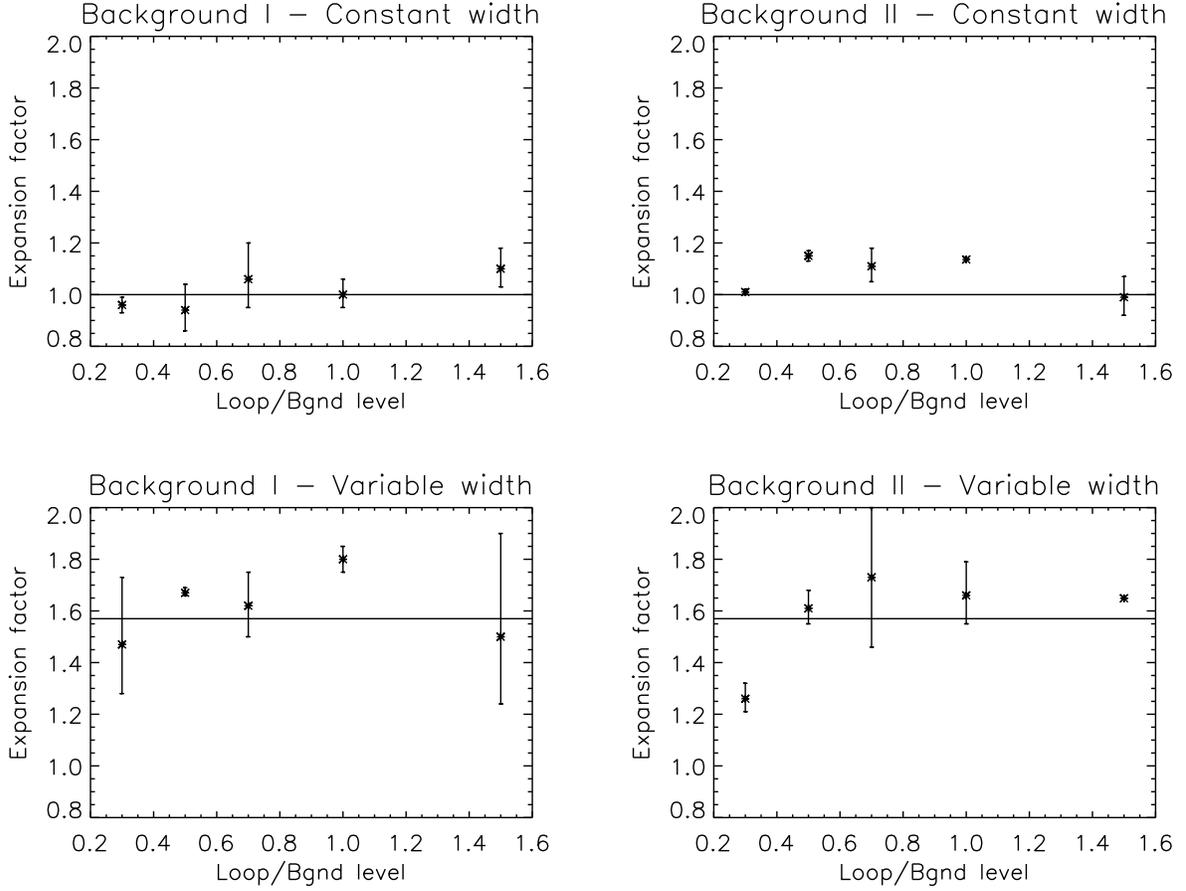}
      \caption{Expansion factor $\Gamma_{m/se}$ versus loop-to-background intensity ratio $\Phi$, for expanding and non-expanding synthetic narrow 
loops on backgrounds I and II (similar to loops in 
panels e-h of Figure~\ref{f4}). The error bars are defined by 
the expansion factors $\Gamma_{m/s}$ and $\Gamma_{m/e}$ (see 
Section~\ref{results-1}). The horizontal line indicates the expansion 
factor of the model.}
         \label{f5}
\end{figure*}

\begin{figure*}   
   \centering
  \hspace{0cm}
\includegraphics[bb= 112 257 459 451,width=8.cm]{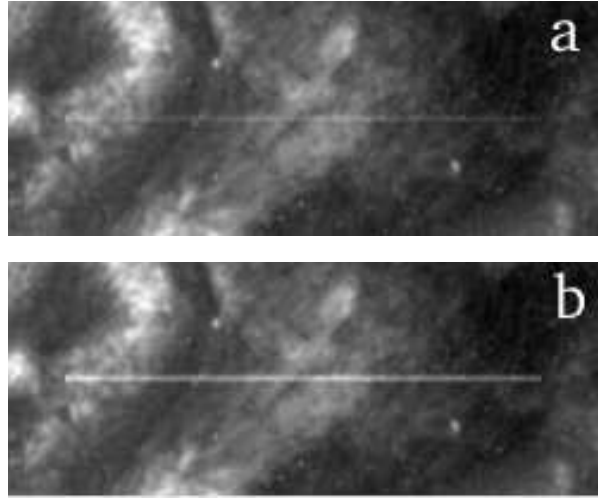}
      \caption{Two examples of unresolved synthetic loops constructed with a constant width of 0.5 pixels. The loops differ only in the loop-to-background 
intensity ratio $\Phi$, which is set to 1 for case (a) and 3 for case (b) 
(see Section~\ref{results-1}).}
         \label{f6}
\end{figure*}

\begin{figure*}   
   \centering
  \hspace{0cm}
\includegraphics[bb= 60 90 510 790,width=12cm]{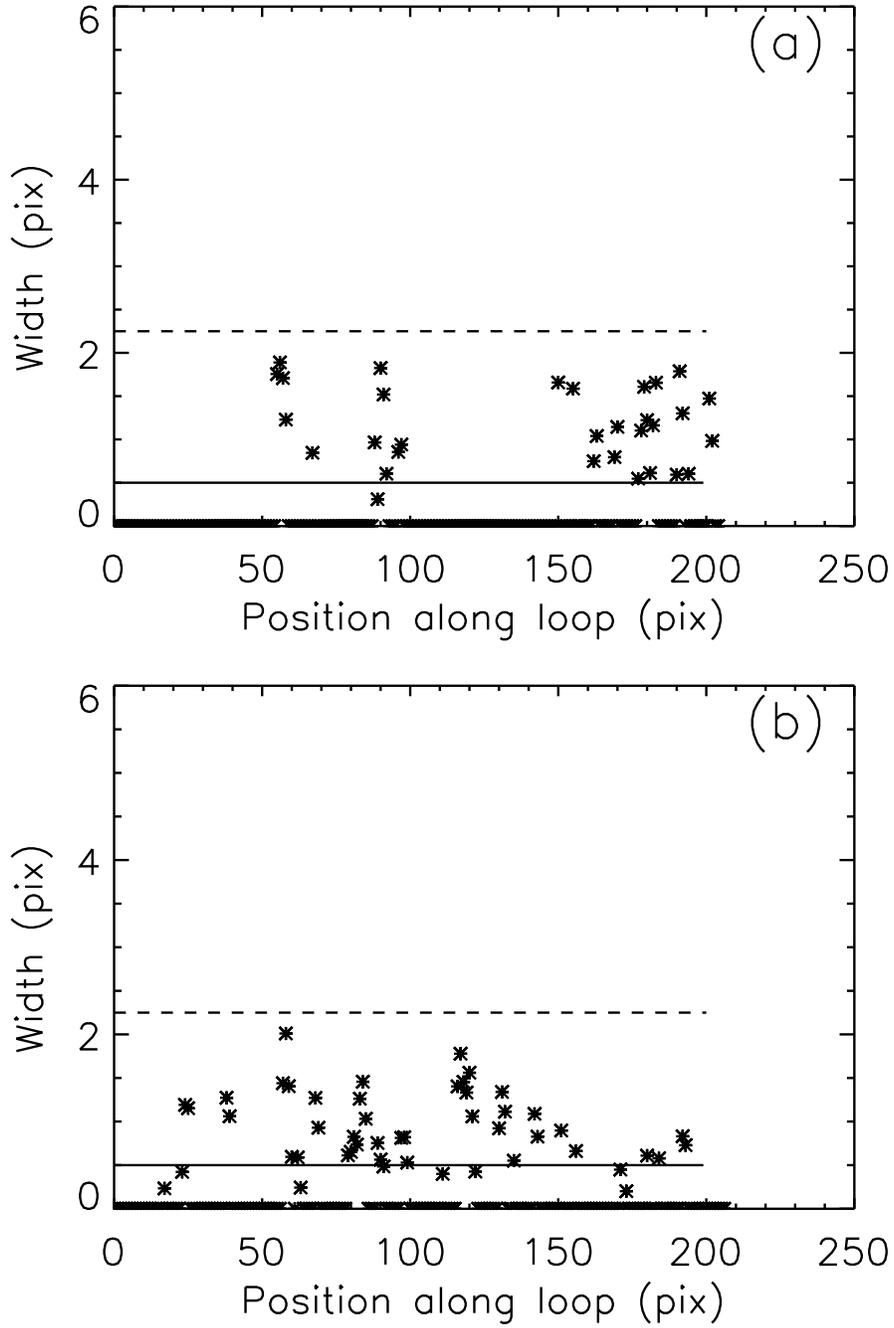}
      \caption{Width measurements for the two synthetic loops in 
Figure~\ref{f6}, corrected for the instrument resolution.  The 
solid line indicates the actual model loop width, and the dashed line 
indicates the PSF full width at half maximum.}
         \label{f7}
\end{figure*}

It is interesting to compare the synthetic images in Figure~2 in 
DeForest's article with our Figure~\ref{f3}. There, he claims that
synthetic loops made from a single unresolved thread of constant 
width and from two diverging threads are indistinguishable 
from each other and from actual 
\textit{TRACE} loops. In our Figure~\ref{f3} it is also very difficult,
by eyeball, to determine which loops have expanding widths or constant 
widths. However, the plots in Figure~\ref{f4} show that a careful 
examination through a quantitative measurement provides the answer. 

One of DeForest's main arguments is that it is difficult to measure the 
width of features that are at or near the instrument resolution due to 
effects such as the smearing from the telescope, pixilation from the detector, 
and the presence of background emission.   We agree, but these claims need 
to be quantified.  It is not sufficient to make eye-ball comparisons of 
features.  Quantitative measures must be used.  We have adopted the 
standard deviation of the loop's cross-axis intensity profile as one 
such measure.  We have been very careful in our work to indicate when the 
measurements are reliable and when they are not.  Measured widths that are 
very close to the instrument resolution have very large error bars that we 
show (see LKD06) and that we take into account.

We have paid particularly careful attention to the effects of the 
combined PSF, which accounts for both smearing and pixilation.  
DeForest is correct that measurements of very thin 
features depend critically on the PSF.  We have therefore adopted a 
conservative value for the PSF width that is greater than the estimates 
determined by the instrument teams and others.  Furthermore, features 
as narrow as our assumed PSF are routinely observed, which would not 
be possible if the actual PSF were wider.

DeForest is also correct that background emission can be important and 
may lead to spurious results.  It is therefore vital to subtract the 
background before making measurements, as we have done.  
In LKD06, we have avoided 
loops where the background is especially bright or complicated.  We 
attempted in our earlier studies to estimate the uncertainties associated 
with imperfect background subtraction, but this was not as careful as 
our treatment of resolution effects.  The main purpose of this paper 
was to rigorously evaluate the effects of background on the measurement 
of loop widths. 

Regarding the importance of quantitative measurements versus 
visual inspection, we concur with DeForest that the visual determination 
of the edge of a feature is subjective and largely based on the intensity 
gradient across the feature. This can lead to erroneous conclusions 
about width variations if there is a systematic variation of intensity 
along the feature, such as decreasing intensity with height. Our 
quantitative measure of width based on the standard deviation of the 
intensity profile is by construction moderating such bias. The 
positive correlation found between the loop width and the maximum intensity
(top panels in Figure~\ref{f10}) could be a remnant of this effect or 
an intrinsic property of the loops. 

DeForest correctly points out that, with optically-thin coronal emission, 
the observed intensity scale height of a hydrostatic structure is larger 
for an expanding loop than for a constant cross section loop, especially 
if the loop is unresolved.  In fact, for the 1-2 MK model examples he shows 
(Figures 5 and 6), the intensity actually increases with height by a 
factors of 2-3 to a maximum brightness at altitudes near $7\times10^9$ cm).
Whether actual TRACE loops have this property is unknown and 
should be investigated.  The variation of temperature with height 
combined with the transmission properties of the filter used will 
complicate the interpretation. 

We note that the observation of super-hydrostatic scale heights is 
different from the observation of excess densities in TRACE loops. 
For most TRACE loops, the density inferred from the observed emission 
measure and diameter is much larger than that expected from static equilibrium 
theory, given the observed temperature and loop length (Aschwanden et al. 
2001, Winebarger et al. 2003).  DeForest's idea of unresolved loops would 
make this discrepancy even worse, since a higher density is required to 
produce the same emission measure from a smaller volume.

The loops identified and measured by DeForest are qualitatively 
much different from the loops identified and measured in our studies.
We chose cases that are not obviously composed of a few resolved or 
quasi-resolved strands (although we believe that our loops may be composed 
of large numbers of elemental strands that are far below the resolution 
limit). The only one of his loops with no apparent internal structure 
(Loop 6 in his 
Figure 8) would have not been selected by us, because it is barely 
discernable above the background. On the other hand, some of the thinner 
structures within DeForest's loop bundles (e.g., at the bottom edge 
of his Loop 3) are not unlike the loops we have investigated. In this 
regard, we must clarify a comment attributed to one of us at the end of 
Section 5 in his paper.  We suggest that researchers seeking to study 
monolithic-looking loops will tend to select cases that are only a few 
resolution elements across.  Significantly wider loops (e.g, all except 
Loop 6 in DeForest's sample) usually show evidence of internal 
structure and will be rejected.

We agree fully with DeForest that collections of loops (loop bundles)  
expand appreciably with height.  However, we stand by our claim that  
individual loops that are clearly discernable within a bundle have a 
much more uniform width.  This is not an artifact of the resolution.  
A hare and hounds exercise, as currently planned, is one useful way 
to clarify any remaining differences of opinion.


An important topic of the present study has been the analysis of
how the properties of the background affect the loop width determination. 
We searched for correlations between the width and: the loop intrinsic 
intensity, the background intensity to loop intensity ratios, and the 
absolute value of the background difference. The background intensity 
is computed as the average between the pixels at both sides of the loop, 
which is used for the estimation of the actual loop background, while 
the background difference is the difference between those pixels.
We found a direct correlation between the width and the maximum
intensity of the loop profile (see Figure~\ref{f10}, upper panels). 
This is probably due to the fact that we tend to miss the ``tails'' 
of the loop 
profile at positions where the loop is less intense with respect to 
the background, and therefore, the measured profile tends to be 
narrower. This is confirmed by the inverse correlation found between 
the width and the ratio of background intensity to loop intensity 
(see Figure~\ref{f10}, bottom panels). It can be seen from the plots 
that the width tends to be abnormally narrower, and the points more
disperse, for larger background to loop intensity ratios.

We found no evidence of correlation between the width and the
the background difference. This shows that the background gradients 
are less important in the determination of the width than the 
background relative intensity. We stress, however, that this does
not affect our ability to determine the global expansion properties 
of loops, and that despite of the background contribution we are 
readily able to distinguish constant width loops from loops that 
expand as predicted from simple force-free magnetic models.
 
The results presented here are extremely intriguing and provide clues 
and new questions to guide future investigations. However, it is expected  
that definitive answers will come from improved observations using new 
generations of solar instruments with higher resolution.  
 

\acknowledgements

We acknowledge the Transition Region and Coronal Explorer 
(\textit{TRACE}) team. We wish to thank Craig DeForest for
fruitful discussions about the nature of observed loops.
We also thank our anonymous referee for his/her valuable suggestions
and comments. The authors acknowledge financial support from 
CNRS (France) and CONICET (Argentina) through their cooperative 
science program (N$^0$ 20326). 
MLF thanks the Secretary 
of Science and Technology of Argentina, through its RAICES program, 
for travel support. This work was partially funded by NASA and the 
Office of Naval Research. 


\clearpage


\begin{table}
\caption{Expansion factors $\Gamma_{m/se}$ (Equation~\ref{gamma}) 
for the synthetic loops shown in Figures~\ref{f3}
and~\ref{f4} (see detailed explanation in Section~\ref{results-1}).}
\label{table1}
\vspace{0.5cm}
$\begin{array}{lccc}
$Synthetic loop$             & $Imposed$ & $Background I$ & $Background
II$ \\
\hline 
$Const. width (4 pix)$       & 1    & 0.85 \pm .02 & 0.95 \pm .04  \\
$Variable width (2.5-5 pix)$ & 1.57 & 1.38  \pm .25  & 1.76 \pm .01  \\
$Const. width (3 pix)$       & 1    & 0.82 \pm .05  & 1.03 \pm .03  \\
$Variable width (2-4 pix)$   & 1.57 & 1.59 \pm .28  & 2.11  \pm .20  \\
\hline
\end{array}$
\end{table}


\clearpage

\clearpage
\thispagestyle{empty}
\setlength{\voffset}{-20mm}
\clearpage
\setlength{\voffset}{0mm}

\begin{figure*}   
   \centering
\includegraphics[bb= 90 138 584 691,width=15cm]{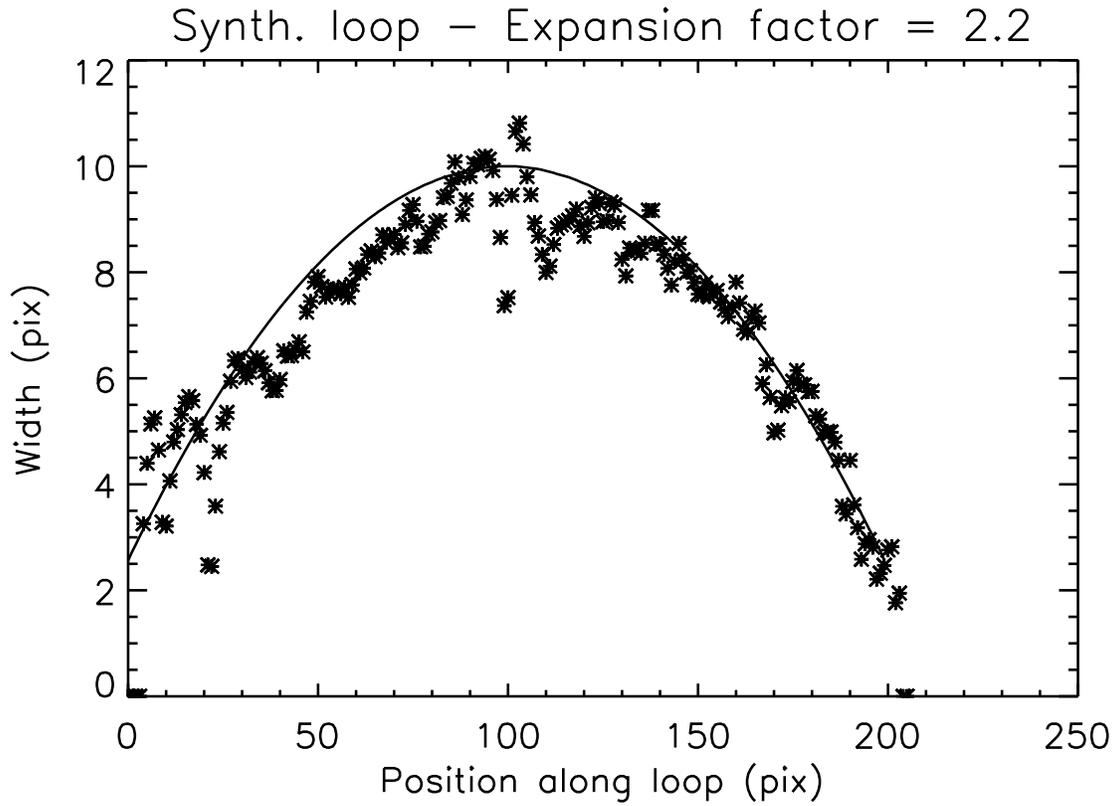}
      \caption{Width versus position along a synthetic loop with 
a model expansion factor of 2.2. The width has been corrected for 
instrument resolution.}
         \label{f8}
\end{figure*} 

\begin{figure*}   
   \centering
\includegraphics[bb= 55 140 480 740,width=15cm]{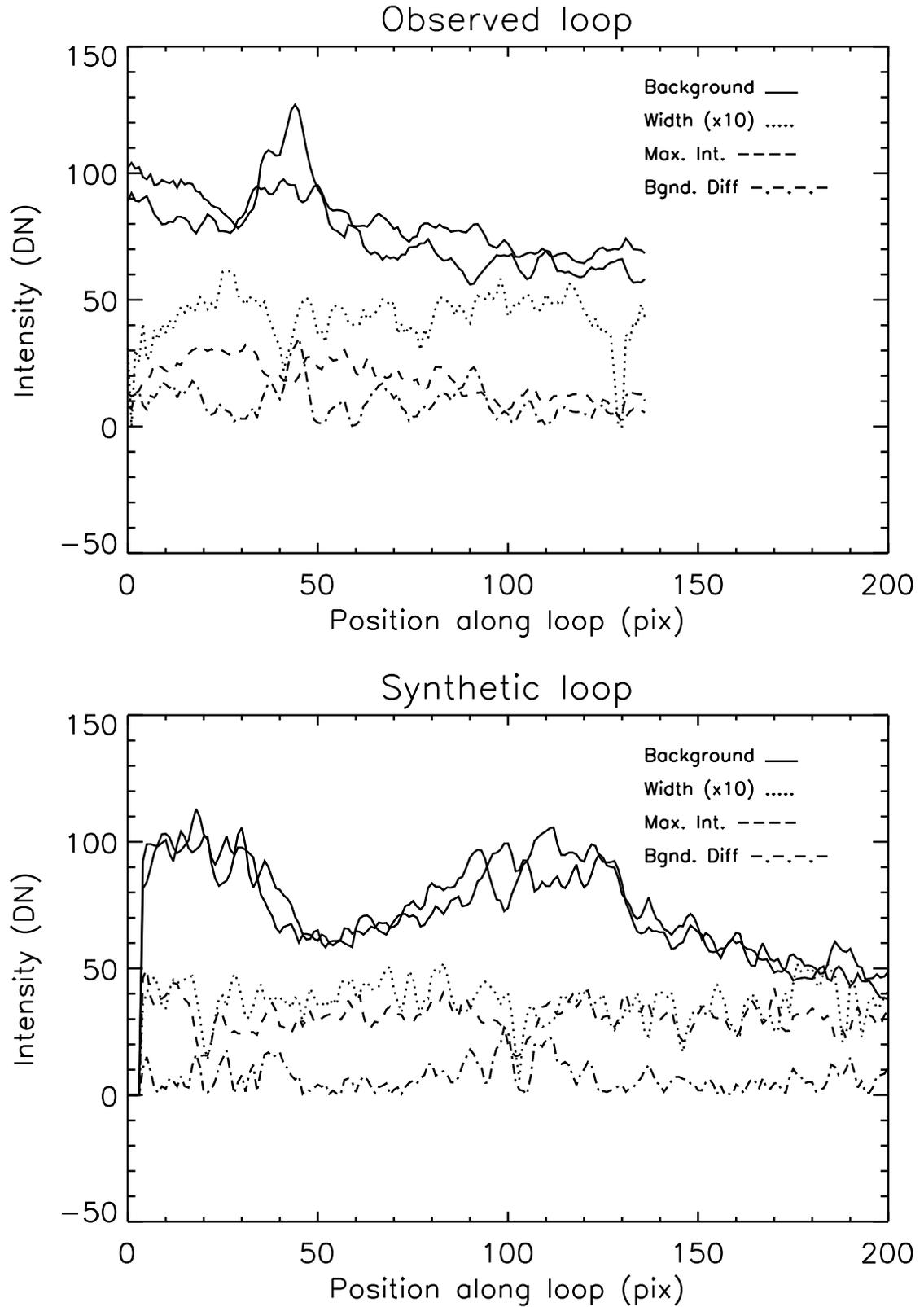}
      \caption{Different loop and background properties versus position 
along the loop. Top panel: example loop from
Figure~\ref{f1}; botton panel: synthetic loop from Figure~\ref{f3}, panel
a). For a detailed description see Section~\ref{results-2}. }
         \label{f9}
\end{figure*} 

\clearpage
\thispagestyle{empty}
\setlength{\voffset}{-20mm}
\begin{figure*}   
   \centering
  \hspace{0cm}
\includegraphics[bb= 65 245 570 640,width=16.5cm]{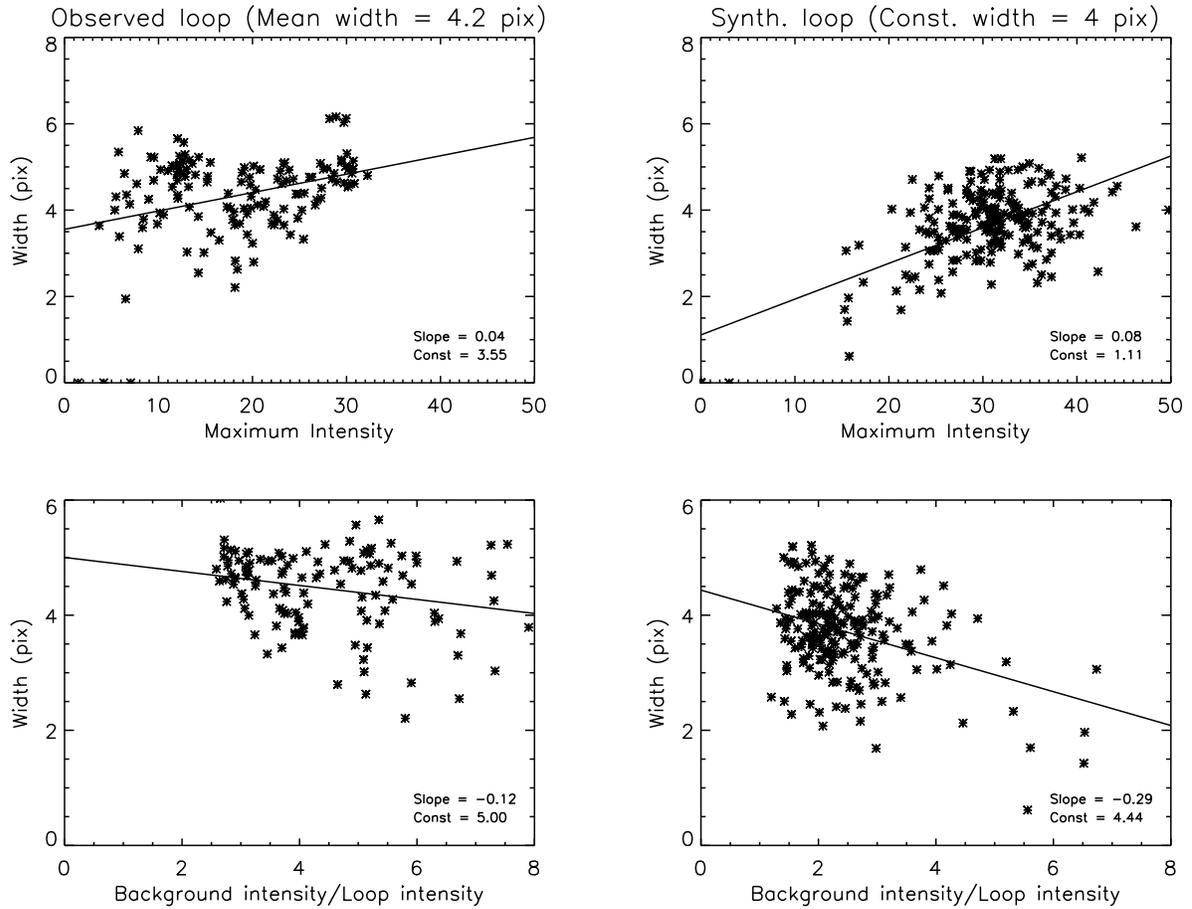}
      \caption{ Scatter plots of measured quantities for 
the observed loop in Figure~\ref{f1} (left column) and for the 
synthetic loop of Figures~\ref{f3} and ~\ref{f4}, panels (a) (right column). 
Top: width versus on-axis loop intensity. 
Bottom: width versus background intensity to loop intensity ratio
(see Section~\ref{results-2}). Continuous lines correspond to least-squares
fits of the data. Slopes and intercepts are given in the respective panels.}  
         \label{f10}
\end{figure*} 

\clearpage
\thispagestyle{empty}
\setlength{\voffset}{-20mm}
\begin{figure*}   
   \centering
  \hspace{0cm}
\includegraphics[bb= 65 245 570 640,width=16.5cm]{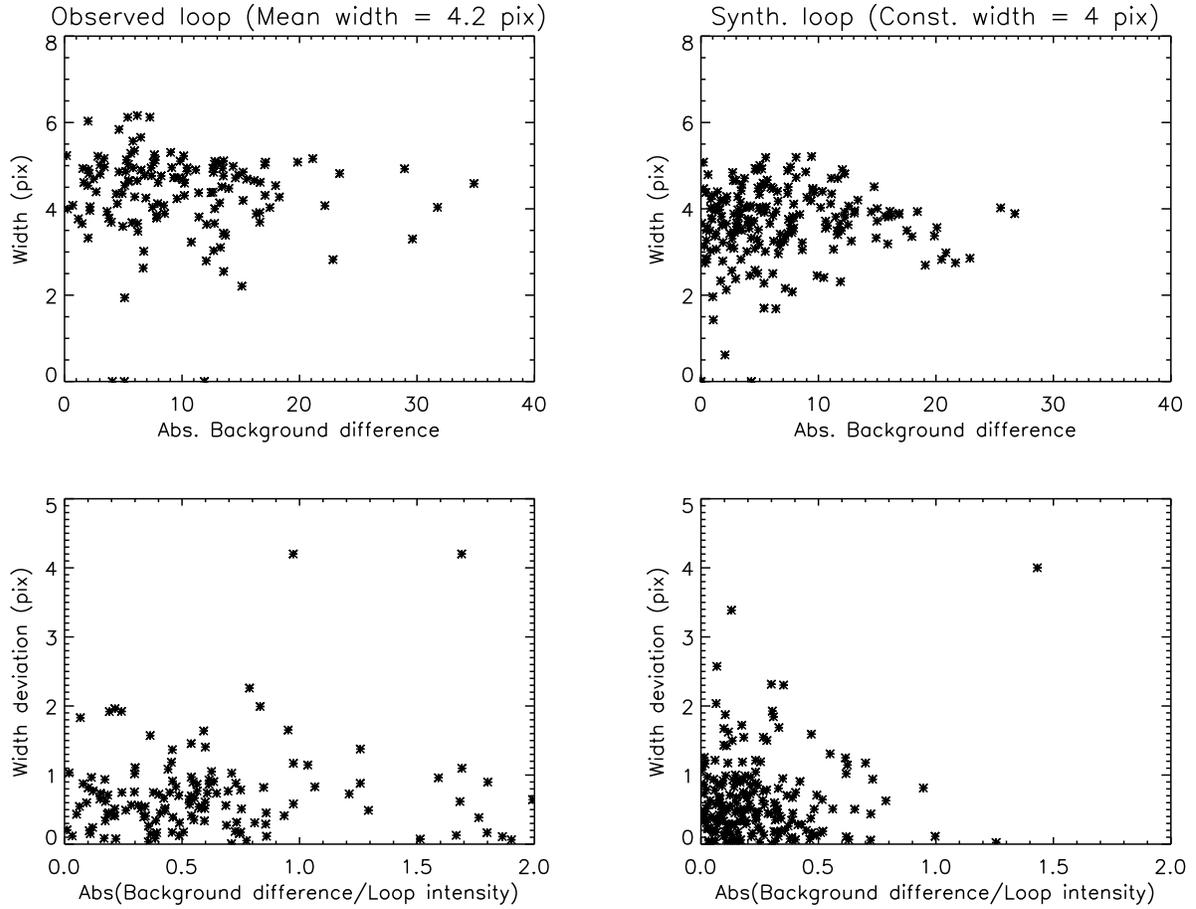}
      \caption{ Scatter plots of measured quantities for 
the observed loop in Figure~\ref{f1} (left column) and for the 
synthetic loop of Figures~\ref{f3} and ~\ref{f4}, panels (a) (right column). 
Top: width versus absolute value of the background intensity difference 
across the loop. 
Bottom left: width deviation from the mean versus  
the background intensity difference normalized by the loop intensity 
(see Section~\ref{results-2}).  
Bottom right: same kind of plot, but the 
deviation is relative to the model width.}
         \label{f11}
\end{figure*}

\end{document}